\documentclass[10pt,journal,compsoc]{IEEEtran}
\pdfoutput=1

\ifCLASSOPTIONcompsoc
  \usepackage[nocompress]{cite}
\else
  \usepackage{cite}
\fi

\usepackage{orcidlink}
\usepackage{amsmath}
\usepackage{amsthm}
\usepackage{amsfonts}
\usepackage{multirow}
\usepackage{multicol}
\usepackage{listings}
\AtBeginDocument{\DeclareCaptionSubType{lstlisting}}
\usepackage{adjustbox}
\usepackage{caption,subcaption}
\usepackage{stfloats}
\usepackage{commath}
\usepackage{booktabs}
\theoremstyle{definition}
\newtheorem{definition}{Definition}

\usepackage{bp-listings}
\usepackage{xspace}
\newcommand{\eg}{\emph{e.g.,}\xspace}
\newcommand{\ie}{\emph{i.e.,}\xspace}

\usepackage{hyperref}
\usepackage{enumitem}

\usepackage{pgf-lc}

\usepackage{array}
\newcolumntype{R}[1]{>{\raggedleft\let\newline\\\arraybackslash\hspace{0pt}}m{#1}}
\newcolumntype{C}[1]{>{\centering\let\newline\\\arraybackslash\hspace{0pt}}m{#1}}

\begin{document}

\title{What Petri Nets Oblige Us to Say\\
{\Large Comparing Approaches for Behavior Composition}}

\author{Achiya~Elyasaf\orcidlink{0000-0002-4009-5353}, Tom~Yaacov\orcidlink{0000-0002-0565-6506}, and Gera~Weiss\orcidlink{0000-0002-5832-8768}
\IEEEcompsocitemizethanks{\IEEEcompsocthanksitem All authors are with Ben-Gurion University of the of the Negev, Israel. E-mail: \{achiya,geraw\}@bgu.ac.il, tomya@post.bgu.ac.il}}

\IEEEtitleabstractindextext{%
\begin{abstract}
We identify and demonstrate a weakness of Petri Nets (PN) in specifying composite behavior of reactive systems. Specifically, we show how, when specifying multiple requirements in one PN model, modelers are obliged to specify mechanisms for combining these requirements. This yields, in many cases, over-specification and incorrect models. We demonstrate how some execution paths are missed, and some are generated unintentionally. 
To support this claim, we analyze PN models from the literature, identify the combination mechanisms, and demonstrate their effect on the correctness of the model.
To address this problem, we propose to model the system behavior using behavioral programming (BP), a software development and modeling paradigm designed for seamless integration of independent requirements. 
Specifically, we demonstrate how the semantics of BP, which define how to interweave scenarios into a single model, allow avoiding the over-specification. 
Additionally, while BP maintains the same mathematical properties as PN, it provides means for changing the model dynamically, thus increasing the agility of the specification. 
We compare BP and PN in quantitative and qualitative measures by analyzing the models, their generated execution paths, and the specification process.
Finally, while BP is supported by tools that allow for applying formal methods and reasoning techniques to the model, it lacks the legacy of PN tools and algorithms. To address this issue, we propose semantics and a tool for translating BP models to PN and vice versa.
\end{abstract}

}

\maketitle
\IEEEdisplaynontitleabstractindextext

%
\IEEEpeerreviewmaketitle

\ifCLASSOPTIONcompsoc
\IEEEraisesectionheading{\section{Introduction}\label{sec:Introduction}}
\else
\section{Introduction}
\label{sec:Introduction}
\fi
\IEEEPARstart{T}{he} \textit{linguistic relativity} hypothesis says that the languages we speak influence our worldview or cognition. While early linguistics believed that language \textit{determines} thought, it is now commonly accepted that language influences only certain cognitive processes in non-trivial ways~\cite{pae2020linguistic}. Deutscher~\cite{deutscher2010through} for example, says, \emph{``when your language routinely obliges you to specify certain types of information, it forces you to be attentive to certain details in the world and to certain aspects of experience that speakers of other languages may not be required to think about all the time.''}

The linguistic-relativity hypothesis has been a guiding principle for computer languages, from early to modern ones, that were designed to direct programmers to change and adapt their thinking to the way machines ``think''. There are many examples: Iverson argued that notations aid in thinking about computer algorithms~\cite{iverson2007notation}, Matz says that he was inspired by this hypothesis when creating the Ruby language~\cite{Matz2003Ruby}, and many more~\cite{moyne1975relevance, chen2018linguistic}. While linguistics researchers have moved to ``softer'' versions of the theorem, in software engineering, to the best of our knowledge, computer-languages researchers are still guided by the early version of the theorem, with one exempt, as discussed below. 

In this work, we follow Deutcher and demonstrate how the \emph{Petri net} (PN) language routinely obliges users to specify things that they do not wish to specify, resulting in unexpected complications and even incorrect specifications. Specifically, we show how the attempt to specify multiple requirements in one PN model forces users to specify also mechanisms for combining these requirements, resulting in over-specification and possibly incorrect models, as some execution paths are missed and some are generated unintentionally.

\textit{Petri net} (PN) is a modeling language with formal semantics that allow for both executing the model and analyzing it. The formal semantics differentiate PN from other process and behavioral modeling languages, such as activity and sequence diagrams. The ability to synthesize the model into working software and analyze it makes it commonly used for modeling and programming discrete event systems (DES) -- dynamic systems with discrete, potentially infinite, state space. A comprehensive introduction to PN can be found in~\cite{reisig2012petri}. 

In~\autoref{sec:related-work} we survey behavior-composition approaches for PN that have been proposed over the years. Nevertheless, these approaches require modelers to consider all the mutual dependencies directly. As we demonstrate in this paper, this may not be feasible in some cases.

To support our claim on PN, we begin with an analysis of a known DES benchmark, called \textit{level crossing}, and demonstrate the inaccuracy of several PN models for this benchmark. To address this problem, we propose a different modeling and programming paradigm, called \textit{behavioral programming} (BP), that allows for a direct specification, execution, and verification of requirements. Like PN, BP is supported by tools for applying formal methods and reasoning techniques to the model. Nevertheless, to keep the legacy of PN tools and algorithms, we propose tools for translating BP models to PN and vice versa. As we will show, our approach has the following advantages:

\begin{itemize}
\item It supports a modular specification approach where each module isolates a specific aspect of the system behavior.

\item It allows modelers to specify the behavior only, exempting them from specifying mechanisms of combining the behaviors.

\item It allows for applying formal methods and reasoning techniques for analyzing and verifying different properties of the system behavior, such as reachability, liveness, boundedness, etc. 

\item These algorithms can be executed in a compositional way, thus handling large-scale programs.

\item We present transnational semantics from the BP model to the PN and vice versa.

\item This translation supports current PN-based practices and algorithms.
\end{itemize}

Furthermore, to avoid the necessity of verifying properties of both the BP model and the PN model, we propose an algorithm and a tool for comparing the two models and testing their equivalency. Thus, our approach allows PN modelers to verify the requirements' correctness and the alignment between the requirements and their implementation.

PN has many extensions and variations. In this paper, we compare BP to the basic PN formalism and arguably the most familiar one. Nevertheless, we discuss some of these extensions in \autoref{sec:related-work}.

The rest of the paper continues as follows. \autoref{sec:bp} gives a short primer on behavioral programming, followed by a general description of the level-crossing benchmark in \autoref{sec:lc-benchmark}. We model the benchmark requirements with BP in \autoref{sec:modeling-bp} and with PN in \autoref{sec:modeling-pn}. In \autoref{sec:comparing-bp-and-pn}, we provide an algorithm for comparing the two models and use it for demonstrating how the mechanism specification in PN results in incorrect behavior. In \autoref{sec:semantics}, we provide translational semantics between the two models. In \autoref{sec:results}, we complete our analysis with more PN models and quantitative comparison between BP and PN. We conclude the paper with a survey of related work (\autoref{sec:related-work}) and a short discussion (\autoref{sec:discussion}).

\section{A Short Primer on Behavioral Programming}
\label{sec:bp}
The behavioral programming paradigm focuses on constructing reactive systems incrementally from their expected behaviors~\cite{harel2012behavioral,Elyasaf2020COBP}. When creating a system using BP, developers specify a set of scenarios that may, must, or must not happen. Each scenario is a simple sequential thread of execution and is thus called a \emph{b-thread}. B-threads are typically aligned with system requirements, such as ``train may not enter when barriers are up". The set of b-threads in a model is called a behavioral program (\emph{b-program}). During runtime, an application-agnostic execution engine interweaves all b-threads participating in a b-program, yielding a complex behavior consistent with all said b-threads. As we will show, this execution engine exempts the modelers from specifying how the requirements interact.

BP is interesting from the linguistic-relativity perspective. Instead of directing its users to a particular way of thinking, the main design goal of the paradigm is precisely the opposite. It aims to enable modelers to specify reactive systems' behavior in a natural and intuitive manner that is aligned with how they perceive the system requirements. To address this goal, several extensions to the paradigm have been proposed to improve this alignment and remove the necessity of specifying mechanisms~\cite{marron2012decentralized,elyasaf2018context,Elyasaf2020COBP}. Also, user studies measured the naturalness and intuitiveness of the paradigm, compared to other paradigms~\cite{Alexandron2014SBPa, Harel2009TeachingVisualFormalisms}. 

Previous demonstrations of BP include a showcase of a fully functional nano-satellite~\cite{bar2019scenario}, robotic controllers~\cite{elyasaf2019using, katz2019fly}, a reactive IoT building~\cite{elyasaf2018context}, a development tool with an integrated model-checking tool~\cite{bar2018bpjs}, and more. Research results on BP cover, among others, model-checking~\cite{harel2011model}, compositional verification~\cite{harel2013relaxing}, runtime look-ahead~\cite{harel2002smart}, synthesis~\cite{kugler2011synthesizing} interactive analysis of unrealizable specification~\cite{maoz2013counter}.
 
Harel, Marron, and Weiss~\cite{harel2010programming} proposed a simple protocol for b-thread synchronization, as follows. The protocol consists of each b-thread submitting a statement before selecting each event that the b-program produces. The statement declares which events the b-thread requests, which events it waits for (but does not requests), and which events it blocks (forbids from happening). After submitting the statement, the b-thread pauses. When all b-threads have submitted their statements, we say that the b-program has reached a \emph{synchronization point}. Then, a central event arbiter selects a single event that was requested and was not blocked. Given this event, the arbiter resumes all b-threads that requested or waited for that event. The rest of the b-threads remain paused, and their current statements are used in the next synchronization point. 

To make these concepts more concrete, we now turn to a tutorial example of a simple b-program. The example presented in this section is an adaptation of one of the first demonstration programs presented in~\cite{harel2010programming} (the hot/cold example).
For convenience and succinctness of the specification, the b-programs in this paper are written using BPjs --- an environment for running behavioral programs written in JavaScript~\cite{bar2018bpjs}. While the b-programs specification may be considered programming rather than modeling, the same program can be specified using diagrammatic implementations of the BP paradigm, including live-sequence charts~\cite{harel2003come} and Blockly~\cite{marron2012decentralized}. Moreover, b-programs can be translated to PN models and vice versa, as described in \autoref{sec:semantics}. 

\textit{The example:} Consider a system with the following requirements:
\begin{enumerate}
\item When the system loads, do `A' three times.
\item When the system loads, do `B' three times.
\end{enumerate}
\autoref{lst:AB-base} shows a b-program (a set of b-threads) that fulfills these requirements. It consists of two b-threads, added at the program start-up. One b-thread, namely \texttt{Do-A}, is responsible for fulfilling requirement \#1, and the second b-thread, namely \texttt{Do-B}, fulfills requirement \#2.

\begin{lstlisting}[
  float=t,
  xleftmargin=.02\columnwidth, xrightmargin=.02\columnwidth,
  numbers=none,
  basicstyle=\footnotesize\ttfamily,
  label={lst:AB-base},
  caption={A b-program that do `A' and `B' three times each. The order between `A' and `B' events is arbitrary.}
]
bthread("Do-A", function() {
  sync({ request: A })
  sync({ request: A })
  sync({ request: A })
})

bthread("Do-B", function() {
  sync({ request: B })
  sync({ request: B })
  sync({ request: B })
})
\end{lstlisting}

The program's structure is aligned with the system requirements. It has a single b-thread for each requirement, and it does not dictate the order in which actions are performed (\eg the following runs are possible: AABBAB, or ABABAB, etc.). This is in contrast to, say, a single-threaded JavaScript program that must dictate exactly when each action should be performed. Thus, traditional programming paradigms are prone to over-specification, while behavioral programming avoids it.

While a specific order of actions was not originally required, this behavior may represent a problem in some cases. Consider, for example, an additional requirement that the user detected after running the initial version of the system:
\begin{enumerate}
\setcounter{enumi}{2}
\item Two actions of the same type cannot be executed consecutively.
\end{enumerate}
While we may add a condition before requesting `A' and `B', the BP paradigm encourages us to add a new b-thread for each new requirement. Thus we add a b-thread, called \texttt{Interleave}, presented in \autoref{lst:AB-interleave}.

\begin{lstlisting}[
  style=BPjs,
%   float=thpb,
  xleftmargin=.02\columnwidth, xrightmargin=.02\columnwidth,
  numbers=none,
  basicstyle=\footnotesize\ttfamily,
  label={lst:AB-interleave},
  caption={A b-thread that ensures that two actions of the same type cannot be executed consecutively, by blocking and additional request of `A' until the `B' is performed, and vice-versa.},
]
bthread("Interleave", function() {
  while(true) {
    sync({ waitFor: B, 
             block: A })
             
    sync({ waitFor: A, 
             block: B })
  }
})
\end{lstlisting}

The \texttt{Interleave} b-thread ensures that there are no repetitions. It does so by forcing an interleaved execution of the performed actions ---  `A' is blocked until `B' is executed, and then `B' is blocked until `A' is executed. This is done by using the \texttt{waitFor} and \texttt{block} idioms. Note that this b-thread can be added and removed without affecting other b-threads. This is an example of a \emph{purely additive} change, where the system behavior is altered to match a new requirement without affecting the existing behaviors.

Recall the discussion in the introduction regarding Deutcher's concept of languages that oblige people to specify things that they do not wish to specify. A critical reader may suspect that BP obliges users to specify unnecessary information for guiding the execution protocol. To answer this, we note that the BP protocol for composing behaviors is implicitly defined and is not part of the model. This protocol is aligned with an implicit protocol that already exists in requirement documents. Each requirement specifies a single aspect of the behavior, and it does not concern itself with other behaviors, though it is clear that the requirements are related to each other. Thus, the implicit protocol of BP does not force the modeler to specify unnecessary information, only use assumptions that already exist in the requirements.


\section{The level-crossing benchmark}
\label{sec:lc-benchmark}
We now turn to describe the level-crossing benchmark that we will use throughout the following sections. 

The level-crossing (LC) domain was first presented in 1987 by~\cite{leveson1987safety} and modeled with Petri nets (PN). It was later used in various research areas of PN modeling and software safety analysis~\cite{liu2014penda, ghazel2016customizable, mazzeo1997systematic}. Although some of these works modified the original model to pertain features to the relevant study, they all followed the same general behavior of~\cite{leveson1987safety}. 

Levenson and Stolzy~\cite{leveson1987safety} defined the model as a controller for a gate at a railway crossing --- an intersection between a railway line and a road at the same level. The railway line has a sensor that signals the controller whenever the train is approaching, entering, or leaving the crossing zone. Based on the signals, the barriers are raised and lowered, ensuring the safety of the trains, \ie that a train cannot be in the crossing zone while the barriers are up. 

While the system behavior is not explicitly specified as a set of requirements, we have extracted the following requirements as we understand them, and we will later refine them:
\begin{enumerate}[label=\arabic*., ref=Requirement \arabic*]
    \item \label{itm:req-1} The railway sensor system dictates the exact event order: train approaching, entering, and then leaving. Also, there is no overlapping between successive train passages.
    \item \label{itm:req-2} The barriers are lowered when a train is approaching and then raised as soon as possible.
    \item \label{itm:req-3} A train may not enter while barriers are up.
    \item \label{itm:req-4} The barriers may not be raised while a train is in the intersection zone. The intersection zone is the area between the approaching sensor and the leaving sensor.
\end{enumerate}

At system initialization, there is no train at the intersection zone, and the barriers are raised. 

We note that these requirements specify the behavior that the controller should enforce, though they do not specify the implementation details. As we will show, our BP implementation will follow this distinction and keep the alignment between the requirements and the model. However, the PN model will add a mechanism for combining the behaviors that will cause incorrect behavior.


\section{Modeling The Requirements with BP}
\label{sec:modeling-bp}
To emphasize the agility of BP models, we begin with a specification that handles only one railway, and we will later extend this model to support multiple railways and faults. 

Following the principles of BP described in \autoref{sec:bp}, each b-thread in \autoref{lst:railway1} is aligned to a single requirement of the system. 

\begin{lstlisting}[
  float=tbph,
%   xleftmargin=.06\textwidth,
%   xrightmargin=.1\textwidth,
  label={lst:railway1},
  caption={A BP program that specifies the requirements for a single railway. Each b-thread is aligned with a single requirement. An application-agnostic execution engine interweaves these b-threads at runtime, yielding a complex behavior that is consistent with each b-thread, liberating the designer from explicitly specifying the joint model.},
]
bthread("R1: Railway Sensors", function() {
  while(true) {
    sync({ request: Approaching })
    sync({ request: Entering })
    sync({ request: Leaving })
  }
})

bthread("R2: Barriers Dynamics", function() {
  while(true) {
    sync({ waitFor: Approaching })
    sync({ request: Lower })
    sync({ request: Raise })
  }
})

bthread("R3: A train may not enter while " +
   "barriers are up", function() {
  while(true) {
    sync({ waitFor: Lower, block: Entering })
    sync({ waitFor: Raise })
  }
})

bthread("R4: Do not raise barriers while a " +
    "train is in the intersection", function() {
  while(true) {
    sync({ waitFor: Approaching })
    sync({ waitFor: Leaving, block: Raise })
  }
})
\end{lstlisting}

The first requirement, describing the order of the sensor's events, is specified in the first b-thread. It continuously requests to ``approach", ``enter", and ``leave", dictating this specific order. We note that a new cycle can start only if the previous train has left the intersection zone, which is aligned with the requirement of no overlapping between successive train passages.

The second b-thread specifies the second requirement of the barriers behavior. It waits for a train to approach and then requests to lower the barriers. When the barriers are down, it requests to raise them as soon as possible. We note that the two barriers events, Lower and Raise, can only happen consecutively, aligned with the system behavior description. 

\ref{itm:req-3} is specified by the third b-thread, which blocks the train from entering while the barriers are up. The first synchronization point blocks the train from entering before lowering the barriers. The second synchronization point ensures that if the barriers are raised between the approaching and the entering events, then the behavior returns to its initial state.

Finally, the last b-thread specifies \ref{itm:req-4}, blocking the raising of the barriers while there is a train in the intersection zone. 

This model demonstrates some merits of the BP modeling approach. The system was modeled in an incremental and modular manner, where each module is aligned with a single requirement and is unaware of other b-threads. We claim that the resulting modules are readable and comprehensible to all stakeholders.
    
\section{Modeling The System with PN}
\label{sec:modeling-pn}
In this section and in \autoref{sec:comparing-bp-and-pn}, we present three PN models for the LC domain. We begin with the original model from 1987 of Levenson and Stolzy~\cite{leveson1987safety} and continue with the two models of Ghazel and Liu~\cite{ghazel2016customizable} from 2016.
As we demonstrate below, all of these models are incorrect, as some execution paths are missed and some are generated unintentionally.

The original model of~\cite{leveson1987safety} is composed of three types of subsystems: railway traffic, barriers, and a barriers' controller. To comply with the specified behavior of the entire system, the modelers specified a mechanism to combine these subsystems. As we discuss below, this mechanism changes the behavior of the model, causing unpredictable side effects.

\begin{figure}
  \centering
\begin{tikzpicture}
    \pic {railway={1}};
\end{tikzpicture}
  \caption{The PN LC model of~\cite{leveson1987safety} for the railway traffic subsystem.}
  \label{fig:railway_lpn}
\end{figure}
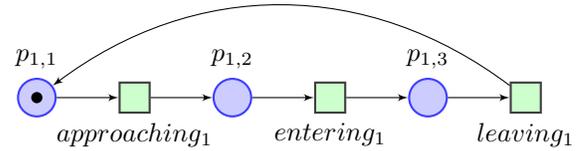

The railway-traffic subsystem (depicted in \autoref{fig:railway_lpn}) specifies the dynamics of the railway using three places and three transitions, corresponding to the sensor's events: $\mathit{approaching}$, $\mathit{entering}$, and $\mathit{leaving}$. The index of these events denotes the railway index, though for now, we have only one.

\begin{figure}
  \centering
\begin{tikzpicture}
    \pic {barrier_display};
\end{tikzpicture}
  \caption{The PN LC model of~\cite{leveson1987safety} for the barriers subsystem.}
  \label{fig:barrier}
\end{figure}
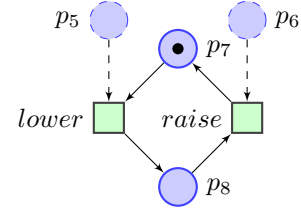

The barriers subsystem (depicted in \autoref{fig:barrier}) has two states --- $\mathit{up}$ and $\mathit{down}$ (marked by $p_7$ and $p_8$ respectively). The barriers passively respond to the commands issued by its controller that we now describe.

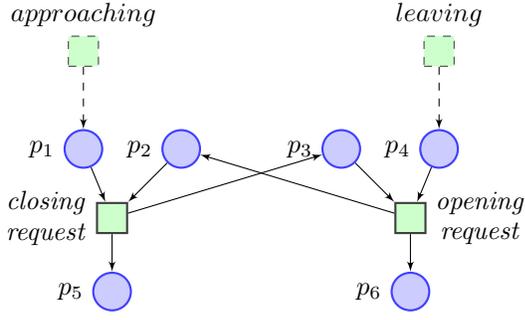
\begin{figure}
  \centering
\begin{tikzpicture}
    \pic {controller_display};
\end{tikzpicture}
  \caption{The PN LC model of~\cite{leveson1987safety} the barrier-controller subsystem.}
  \label{fig:barrier_controller}
\end{figure}

The barriers' controller subsystem (depicted in~\autoref{fig:barrier_controller}) provides an interface between the railway traffic and the barriers subsystems. A $\mathit{closing}$ $\mathit{request}$ is fired when a train approaches, and when a train leaves, an $\mathit{opening}$ $\mathit{request}$ is fired. Note that this subsystem contains two interlocks, $p_2$ and $p_3$, which together make sure that $\mathit{closing}$ $\mathit{request}$ and $\mathit{opening}$ $\mathit{request}$ fire alternatively. In practice, this means that the barriers may be closed if and only if they are open.

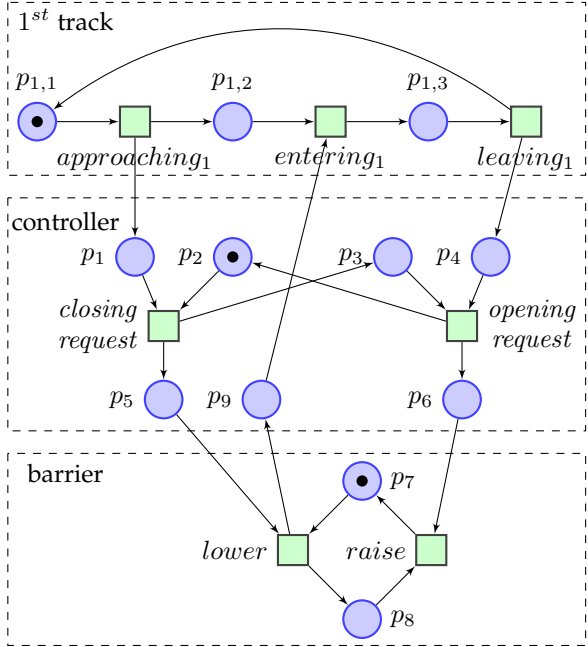
\begin{figure}
  \centering
\begin{tikzpicture}
    \pic {single_track_display};
\end{tikzpicture}
  \caption{The unified PN LC model of~\cite{leveson1987safety}, including the three subsystems and the interlocking mechanism.}
  \label{fig:ghazel-single} 
\end{figure}

Finally, to address \ref{itm:req-3} and forbid the train entrance while the barriers are up, the unified model that integrates the three subsystems (depicted in~\autoref{fig:ghazel-single}) includes an additional interlocking state, $p_9$, and its arcs --- $lower \rightarrow p_9$ and $p_9 \rightarrow entering$. 

We note that the controller events $\mathit{closing}$ $\mathit{request}$ and $\mathit{opening}$ $\mathit{request}$ are not mentioned in the requirements. 
The reason is that Levenson and Stolzy~\cite{leveson1987safety} designed the PN model as a controller where these events act as part of the implementation of the controller. 
We argue, as we show below, that it is better to model the requirements separately of the implementation. 
We mark these implementation events as \emph{helper events}, since they are not required for specifying the system behavior, only for the specific implementation perspective. We show below that the helper events and the interlocking mechanism lead to undesired system behaviors.

\section{Comparing the BP and PN}
\label{sec:comparing-bp-and-pn}
In this section, we compare the BP and the PN models to verify our model's correctness and explicate the differences between the models.

\subsection{Models equivalency}
\label{sec:comparing-bp-and-pn:equivalency}
We begin with a definition of equivalence. Generally, two models are equivalent if they yield the same set of runs, \ie the same sequences of events. However, there is a complication in our case since the PN model requires helper events that are not part of the BP model. For example, the two traces in \autoref{tab:events} are equivalent in system behavior, though they have different events.

\begin{table*}
\caption{Comparing traces of BP and PN. The two traces are equivalent in terms of system behavior, though they have different events.}
\label{tab:events}
\centering
\begin{tabular}{l|lllllll}
\hline \\
BP & Approaching, & & Lower, & Entering, & Leaving, & & Raise  \\ [1ex]
PN & Approaching, & ClosingRequest, & Lower, & Entering, & Leaving, & OpeningRequest, & Raise \\ [1ex]
\hline
\end{tabular}
\end{table*}

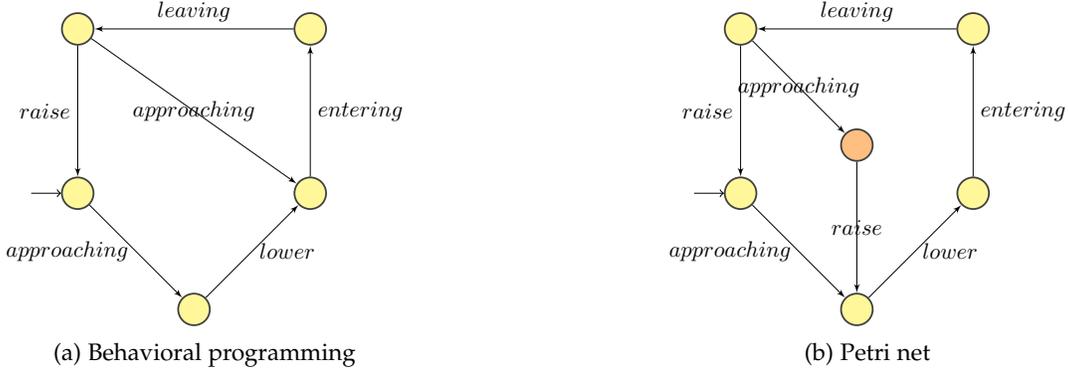
\begin{figure*}
\centering
\begin{subfigure}{.3\textwidth}
  \begin{adjustbox}{width=\linewidth}
  \begin{tikzpicture}
    \pic {bp_automata};
\end{tikzpicture} 
\end{adjustbox}
  \caption{Behavioral programming}
  \label{fig:sub-first}
\end{subfigure}\qquad\qquad\qquad\qquad\qquad
\begin{subfigure}{.3\textwidth}
  \begin{adjustbox}{width=\linewidth}
  \begin{tikzpicture}
    \pic {pn_automata};
\end{tikzpicture}
\end{adjustbox}
  \caption{Petri net}
  \label{fig:sub-second}
\end{subfigure}
\caption{The generated automaton of each of the models for a single track.}
\label{fig:models_automata}
\end{figure*}

As the example shows and noted before, these events are used to synchronize the barriers and the railway events. They are not required to compare the resulted behavior of the two models. Thus, our equivalency definition ignores these events. For completeness, we allow helper events on either side of the comparison.

\begin{definition}
Models $M_1$ and $M_2$ over the event sets $E_1$ and $E_2$, respectively, are equivalent if and only if 
$$\{ \pi_{E_1 \cap E_2}(t) \colon t \in L(M_1)\}  = \{ \pi_{E_1 \cap E_2}(t) \colon t \in L(M_2)\}$$
where $L(M_i)$ is a set of sequences of events, called traces, that model $M_i$ generates,  and $\pi_{E_1 \cap E_2}(t)$ is an operator that removes from a trace $t$ all the events that are not in $E_1 \cap E_2$: 
$$\pi_{E_1 \cap E_2}( t) = \begin{cases} t[0] \pi_{E_1 \cap E_2}(t[1..]) & \text{if } t[0] \in E_1 \cap E_2, \\ \pi_{E_1 \cap E_2}(t[1..]) & \text{otherwise;} \end{cases}$$

\noindent To allow traces with finite length, we also define that $\pi_{E}(\varepsilon)= \varepsilon$ for any $E$. The sequences in the sets $L(M_i)$ can have a finite or infinite length. 
\end{definition}

Using this definition, we denote $M_{BP}$ and $M_{PN}$ as the BP model and the PN model (respectively). Since the trains may infinitely approach, enter, and leave the crossing zone, we use B\"uchi automata to represent the languages  $L(M_{BP})$ and $L(M_{PN})$. 
A B\"uchi automaton consists of a set of states and a transition function, where some states are defined as accepting and some as initial (starting). The automaton accepts input if and only if there is a run over this input that begins at an initial state, and  at least one of the infinitely often occurring states is an accepting state. 

We generate the automata using a depth-first search that traverse the state space of each model, where transitions correspond to events and all states are accepting (depicted in \autoref{fig:models_automata}). Thus, the accepting words of these automata represent the set of possible traces that each model may generate. 

To understand the significance of the difference between the two models, we analyze them using GOAL~\cite{tsay2007goal} --- a graphical tool for manipulating B\"uchi automata and temporal formulas. Our findings show that the resulting language for the BP model is contained in the resulting PN model --- $L_{M_{BP}} \subset L_{M_{PN}}$, meaning that some runs are only possible in the PN model. One example for a word (or a trace) that is accepted only by the PN model is:

\noindent $Approaching\cdot ( Lower\cdot Entering\cdot Leaving\cdot Approaching\cdot\allowbreak Raise)^\omega$

In this case, there are two trains on the same railway, where the second train approaches the intersection zone after the first train leaves while the barriers are down. According to this trace, even though a train is already approaching the barriers, the latter may be raised only to be lowered again immediately after. As we describe in \autoref{sec:multi-track}, \cite{ghazel2016customizable} added the $\mathit{keep}$ $\mathit{down}$ event to avoid this behavior, though it did not completely prevent it and caused other problems to the model. Thus, these redundant barriers actions are not aligned with the system requirements and do not stand to reason. We believe that such behavior is derived from their specific mechanism implementation. While the BP paradigm allows for a direct specification of the requirements as separate modules and their automatic composition, the PN modeling approach obliges the modeler to explicitly specify how the different modules interact.

\subsection{Adjusting our model}
To allow a simple comparison of the models, we now adjust our model to meet this behavior. We start with a redefinition of \ref{itm:req-2}:

``The barriers should be lowered after a train approaches. If the barriers were already lowered, then they should be raised, and immediately lowered again before the train enters the intersection zone".

Granted, this is a strange and tangled requirement, though it describes the observed behavior.

This adjustment requires the modification of the second b-thread. The original b-thread and its modified version are presented in \autoref{lst:bthread2}. Once a train leaves, the controller requests to raise the barriers while waiting for another approaching event, whichever comes first. If the event is approaching, it requests to raise the barriers. Otherwise, it waits for a train to pass again. In addition to this change, the fourth b-thread should be removed, to allow the raise of the barriers after the train approaches.

Given these modifications, the BP and the PN models are now equivalent. 

\begin{figure*}[tbph]
\captionsetup{type=lstlisting}
\begin{sublstlisting}[b]{0.35\linewidth}
\begin{lstlisting}[
    numbers=none,
    xleftmargin=0\textwidth,
    xrightmargin=0\textwidth,
]
bthread("R2: Barriers Dynamics", 
    function() {
  while(true) {
    sync({ waitFor: Approaching })
    sync({ request: Lower })
    sync({ request: Raise })
  }
})
\end{lstlisting}
\caption{The original b-thread}
\label{lst:bthread2:1}
\end{sublstlisting}\hfill
\begin{sublstlisting}[b]{0.6\linewidth}
\begin{lstlisting}[
    numbers=none,
]
bthread("R2*: Modified Barriers Dynamics", function(){
  while (true) {
    sync({ waitFor: Approaching })
    sync({ request: Lower })
    while (true) {
      sync({ waitFor: Leaving })
      let e = sync({ request: Raise, waitFor: Approaching })
      if (e == Raise) {
        sync({ waitFor: Approaching })
        sync({ request: Lower })
      } else {
        sync({ request: Raise, block: Entering })
        sync({ request: Lower })
      }
    }
  }
})
\end{lstlisting}
\caption{The modified b-thread}
\label{lst:bthread2:2}
\end{sublstlisting}
\caption{Adapting the second b-thread to the change in the requirement.}
\label{lst:bthread2}
\end{figure*}

\subsection{Expanding to Multi-Track}
\label{sec:multi-track}
Ghazel and Liu~\cite{ghazel2016customizable} observed this redundant behavior and tried to address it. In addition, they extended the model to support multiple railways. Many have used this extension as a benchmark for this domain~\cite{boussif2016contributions, boussif2017experimental, liu2018time}, and we use it to continue our comparison. \autoref{fig:multi-track} presents the extended PN model of~\cite{ghazel2016customizable} and our extended BP model. 

\newsavebox{\tempbox}
\sbox{\tempbox}{\begin{tikzpicture}
\pic {multi_track_display};
\end{tikzpicture}%
}
\newbox{\mybox}
\begin{lrbox}{\mybox}
\begin{lstlisting}[linewidth=7cm,numbers=none]
function bt(i) {
  bthread("R1 "+i, function() {
    while(true) {
      sync({ request: Approaching(i) })
      sync({ request: Entering(i) })
      sync({ request: Leaving(i) })
    }
  })

  bthread("R3 "+i, function() {
    while(true) {
      sync({ waitFor: Lower, 
               block: Entering(i) })
      sync({ waitFor: Raise })
    }
  })

  // other b-threads are 
  // omitted for brevity. 
}

for (var i = 0; i < n; i++) 
  bt(i)
\end{lstlisting}
\end{lrbox}

\begin{figure*}[tbph]
    \centering
    \subfloat[The extended model of~\cite{ghazel2016customizable}. The changes from the model of~\cite{leveson1987safety} are emphasized.\label{fig:multi-track:pn}]
    {\adjustbox{scale=0.8,valign=b}{\usebox\tempbox}}
    \qquad\qquad\qquad
    \subfloat[The extended BP model.\label{fig:multi-track:bp}]
    {\adjustbox{valign=b}{\usebox\mybox}}
    \caption{Multi-track extensions of the two models. Besides multiplying the railway traffic subsystem, the PN extension consists of additional arcs, tokens, and a transition, as opposed to the BP extension.}
    \label{fig:multi-track}
\end{figure*}

Extending the BP model required only multiplying the b-threads by the number of tracks. Railway-specific events (\ie Approaching, Entering, and Leaving) were assigned with an index while the barriers events remained the same. Since the behavior of each b-thread is valid for both the single version of the system and the multi-track version, we needed no further adaptations to the extended model.

The extended PN model of~\cite{ghazel2016customizable} significantly changed the model and its semantics. To support multi-track, they multiplied the railway traffic subsystem and changed the other two subsystems as follows: two additional arcs were added ($p_6 \rightarrow \mathit{closing}$ $\mathit{request}$ and $raise \rightarrow p_6$), tokens were added, and some arc weights were changed. 

To address the redundant raise-lower behavior, the model adds a helper event, called $\mathit{keep}$ $\mathit{down}$, and its related arcs. These additions allow to keep down the barriers if a train approaches right after another one leaves. Alas, not only that their solution did not solve it in all cases, but it created other problems, as we present in \autoref{sec:results:lc}.

Applying the comparison algorithm of \autoref{sec:modeling-pn} reveals that the extended PN model is, again, not aligned with the requirements and is no longer equivalent to our model. 
In \autoref{sec:results}, we further analyze the differences between these models.




\subsection{Adding Faults}
\label{sec:comparing-bp-and-pn:faults}
In real life, discrete-event systems may have faults. For example, the entering sensor on a railway may be faulted and miss a train entering. Detecting and diagnosing such faults at real-time is of paramount importance in DES modeling and has become an active research area in recent years. The research activity in this area is driven by the needs of many different error-prone domains. When modeling, faults are often added to the basic model that describes the standard system behavior. This may lead to inconsistent system behavior that is misaligned with prior requirements. In \autoref{sec:results:lc}, we give multiple examples for such inconsistencies. Here, we demonstrate how our suggested method can assist modelers in verifying and analyzing the impact of faults on the initial requirements.

\begin{figure}
  \centering
\begin{tikzpicture}
    \pic {multi_track_fault_display};
\end{tikzpicture}
  \caption{The extension of~\cite{ghazel2016customizable} to the PN model that adds faults (new transitions and arcs are emphasized).}
  \label{fig:ghazel-fault} 
\end{figure}
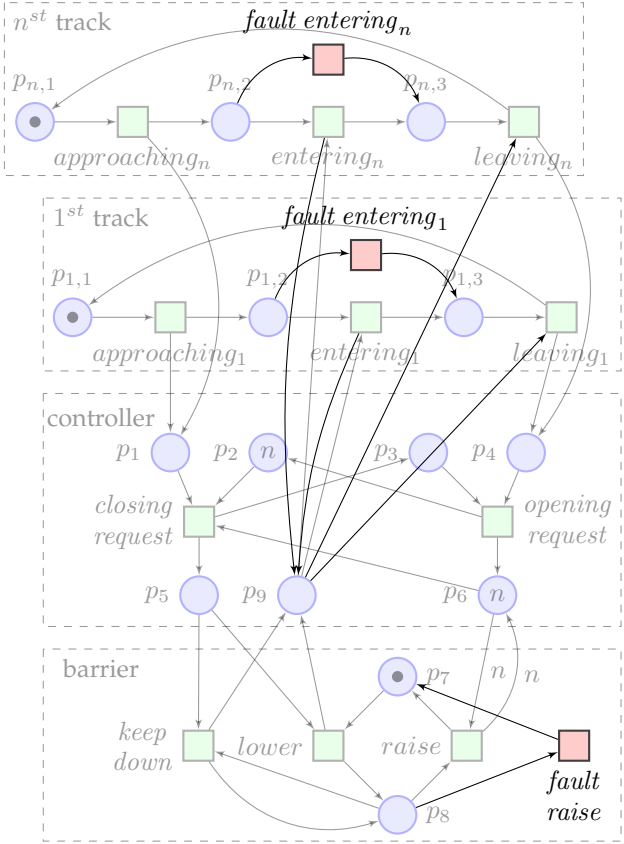

In the PN model detailed in~\cite{ghazel2016customizable}, two classes of faults were added for diagnosis purposes (denoted with red transitions in \autoref{fig:ghazel-fault}). The first one simulates a train-sensing defect and indicates that the train enters the level-crossing zone without triggering the entering sensor. Thus, the train may enter before the barriers are lowered. The second failure indicates a defect of the barriers that result in a premature raising.

A detailed look at their model, reveals that the arcs to and from $p_9$ are not required for modeling the fault transitions. In fact, they were added as part of a mechanism for satisfying the original requirements given the new fault transitions. 
As we show in~\autoref{sec:results:lc}, these arcs fix one behavior and break others, blocking many legit traces that can no longer happen. This example, together with the multi-track extension, demonstrate the drawback of PN for modeling behaviors --- adding new behaviors after the model is ready often requires modifying the mechanism and performing non-trivial adjustments to the model. It obliges modelers to think of all the side effects of these adjustments, something that may not be practical for complex behavior of large-scale systems.

The BP version of the fault transitions is presented in \autoref{lst:fault-bthreads}. The first class of faults, simulating an unobservable train entering, is modeled using $n$ b-threads, one for each railway. Each b-thread waits for a train to approach and then requests a signed entering event, representing the fault. The second b-thread models the second fault class of a premature barriers' raise. The b-thread waits for the barriers to be lowered. It then requests to raise the barriers using a signed event that represents this fault. We note that the fault events in our BP model are similar to the original non-faulted events for both entering and raising respectively and are differentiated by adding a ``flag" to the event data. This setting allows existing b-threads to effect (by blocking) or be affected by (requesting and waiting for) the added fault events without modifying them.

\begin{lstlisting}[
  style=BPjs,
  float=tbph,
  xleftmargin=.02\columnwidth, xrightmargin=.02\columnwidth,
  numbers=none,
  basicstyle=\footnotesize\ttfamily,
  label={lst:fault-bthreads},
  caption={Faults b-threads for the level-crossing benchmark.},
]
let fault = true

bthread("UnobservableEntering_"+i, function() {
  while(true) {
    sync({ waitFor: Approaching(i) })
    sync({ request: Entering(i, fault) })
  }
})

bthread("Premature Raise", function() {
  while(true) {
    sync({ waitFor: Lower })
    sync({ request: Raise(fault) })
  }
})
\end{lstlisting}

\subsection{A Conclusion for the Level-Crossing Domain}
The addition of multiple tracks and fault transitions demonstrate the dynamical and incremental development style of BP. The described b-program is modular because new requirements can be flexibly added as new b-threads. Since modeling often begins without faults, we believe that the ability to implement them without modifying (or even accessing) the existing model is a significant advantage. Furthermore, it maintains the alignment of the model with requirements. System requirements, both old and new, can be represented directly using a BP model. 

At this point, a critical reader may ask whether the BP model is indeed error-free or has some unknown problems. We used the verification tool of BPjs to assert some properties of the model, such as that it has no deadlocks or livelocks. To validate that the model is indeed aligned with the requirements, we sampled the generated traces of the model and checked that they are aligned with the way we perceive the requirements. 

To conclude the different versions of the LC benchmark, we present in \autoref{sec:results:lc} a quantitative comparison between the PN and the BP models for the LC benchmark. This comparison emphasis the effect of the problems in the PN model on the resulted behavior. To the best of our knowledge there are no other references in the literature to these problems in the LC benchmark.

\section{Translational Semantics From PN to BP}
\label{sec:semantics}
BP has tools for executing and analyzing the models (e.g., using formal methods). In addition, there are tools for converting BP models into other formats, such as Z3, GOAL, SPIN, Graphviz, and more. Nevertheless, PN has long-standing successful tools that may be more suitable for different use cases. Thus, we propose an approach for PN modelers to use BP to bridge the gap between system specification and PN implementation. System requirements can be specified directly using BP and implemented using PN. Hence an equivalence between the two models, in such a case, can indicate an alignment between requirements and implementation. This approach ``eases'' the transition for PN modelers and maintains the current advantages of PN. Although the automaton required for the comparison can be computed directly from PN, we now show a viable alternative that is more useful in practice --- a direct translation from PN to BP. Tools based on these semantics can provide a uniform modeling environment where all modeling artifacts are translated to a common language.

Taking advantage of BP's modularity and flexibility, we can implement the dynamics of each place of the PN model as a separate b-thread. Each b-thread maintains the number of tokens in the place using a variable. Based on the number of tokens, it waits-for or blocks a set of events that represents the transitions to/from the place. A translation example for $p_2$ is presented in \autoref{lst:translated-b-thread}. If it has no tokens, it forbids the event ``Closing Request" from taking place while waiting for an ``Opening Request" event, which increases its tokens. If it has some tokens, it waits for both events and increases or decreases its tokens accordingly. The suggested translation is general and can be applied to all places of a PN model. The complete b-program combines all places b-threads and an auxiliary b-thread that requests all possible events at each round (as depicted in \autoref{lst:translated-b-thread2}). This program yields a behavior consistent with the entire PN dynamics.

Based on these semantics, we translated to BP all the PN models in this paper. 
To verify the correctness of each translation, we generated the automaton of the PN model using SNAKES~\cite{pommereau2015snakes}, a Python library for Petri nets. Next, we generated the automaton for the translated (to BP) model. Finally, we verified that the models are equivalence, using the method described in \autoref{sec:comparing-bp-and-pn:equivalency}. The paper's repository (\url{github.com/bThink-BGU/Papers-2022-BP-PN}) contains the source code for all models (PN, translated-PN-to-BP, and BP) and their automata. In \autoref{sec:results:lc}, we empirically evaluate the differences between the BP and the PN/translated models.

\begin{lstlisting}[
  style=BPjs,
  float=thpb,
  xleftmargin=.02\columnwidth, xrightmargin=.02\columnwidth,
  numbers=none,
  basicstyle=\footnotesize\ttfamily,
  label={lst:translated-b-thread},
  caption={$p_2$ translated b-thread},
]
bthread("p_2", function() {
  var tokens = n
  while(true) {
    if(tokens < 1) {
      sync({ waitFor: OpeningRequest(), 
               block: ClosingRequest() })
      tokens += 1
    } else {
      if(sync({ waitFor: [ClosingRequest(),
                          OpeningRequest()] })
        .name === "ClosingRequest" ) {
        tokens -= 1
      } else {
        tokens += 1
      }
    }
  }
})
\end{lstlisting}

\begin{lstlisting}[
  style=BPjs,
  float=thpb,
  xleftmargin=.02\columnwidth, xrightmargin=.02\columnwidth,
  numbers=none,
  basicstyle=\footnotesize\ttfamily,
  label={lst:translated-b-thread2},
  caption={The ``auxiliary'' translated b-thread},
]
bthread("auxiliary", function() {
  while(true)
    sync({ request: 
       [ClosingRequest(), OpeningRequest()] })
})
\end{lstlisting}

Although a translation from BP to PN is not necessary for the context of this paper, it can be easily done. An automaton representing the behavior of the model, such as the one depicted in~\autoref{fig:models_automata}, can be automatically generated (elaborated in \autoref{sec:comparing-bp-and-pn:equivalency}). This automaton can be viewed as a special case of a simple PN with a single token passing from states (or places). 

\section{Results}
\label{sec:results}
The example of the level-crossing benchmark allowed us to demonstrate our claims on PN. The purpose of this section is to further establish our claims in two ways: 1) quantifying the differences between PN and BP, and; 2) demonstrating our claims on other PN models to support our hypothesis that the problem is rooted in the language constructs.

All the code examples in this paper and the data we used for comparing the models can be viewed at \url{github.com/bThink-BGU/Papers-2022-BP-PN}.

\subsection{Empirical Results for the Level-Crossing Benchmark}
\label{sec:results:lc}
To quantify the differences between the two approaches, we computed the state space (\ie reachability graph) of the PN and the BP programs, with and without failures. We ran each program with a varying number of railways. To evaluate the effect of the helper events on the state space, we removed all transitions $s \xrightarrow{e} t$, where $e$ is a helper event, and rewired all incoming transitions of $s$ into $t$. We denote the resulting state space as PN*.

The results are summarized in \autoref{tab:lc-results}. For comparison, we use the BP model without the \ref{itm:req-2}, since it better aligns with the requirements. We implemented the PN model of~\cite{ghazel2016customizable} using the translational semantics presented in~\autoref{sec:semantics}. Notably, for any $n$ with faults, the number of states and transitions of the PN model matches the reported numbers of~\cite{ghazel2016customizable}, thus validating our implementation (they did not provide statistics for other implementations). To make the differences between the models accessible to the reader, in \autoref{tab:lc-reduction}, we present the state and transition reduction between the different models. For all models, the average reduction in states is similar to the average reduction in transitions. From PN to PN*, the model is reduced by almost 50\%; from PN* to BP, the model is reduced by almost 70\%, and finally; from PN to BP, the model is reduced by approximately 80\%. This dramatic reduction allows for applying reasoning techniques and formal methods on larger models compared to PN. Furthermore, the composable structure of BP programs allows for applying compositional verification and compositional formal method techniques~\cite{harel2013relaxing}.

\begin{table*}
\parbox{.5\linewidth}{
\centering
\caption{Statistics on the models' state space, measuring the number of states and transitions. PN* represents the state space of the PN model that was stripped from the helper events.}
\label{tab:lc-results}
\begin{tabular}[b]{cc|rr|rr|rr} 
\toprule
 & \multirow{2}{20px}{\centering with faults} & \multicolumn{2}{c|}{PN} & \multicolumn{2}{c|}{PN*} & \multicolumn{2}{c}{BP} \\
$n$ & & $\abs{S}$ & $\abs{T}$ & $\abs{S}$ & $\abs{T}$ & $\abs{S}$ & $\abs{T}$ \\ 
\midrule
1 &  & 10 & 13 & 6 & 8 & 5 & 6\\
1 & \checkmark & 20 & 43 & 13 & 26 & 13 & 26\\[5px]
2 &  & 83 & 185 & 37 & 87 & 13 & 26\\
2 & \checkmark & 142 & 500 & 85 & 284 & 39 & 126\\[5px]
3 &  & 483 & 1,532 & 168 & 537 & 35 & 101\\
3 & \checkmark & 832 & 4,085 & 441 & 2,007 & 109 & 490\\[5px]
4 &  & 2,434 & 10,110 & 863 & 3,509 & 97 & 372\\
4 & \checkmark & 4,314 & 27,142 & 3,317 & 19,160 & 325 & 1,868\\[5px]
5 &  & 11,304 & 58,327 & 5,798 & 29,065 & 275 & 1,327\\
5 & \checkmark & 20,556 & 157,551 & 15,421 & 107,290 & 975 & 6,822\\
\bottomrule
\end{tabular}
}
\hfill
\parbox{.45\linewidth}{
\centering
\caption{The state-space reduction between the models, in terms of number of states and transitions}
\label{tab:lc-reduction}
\begin{tabular}[b]{lc|cc|cc|cc}
\toprule
 & \multirow{2}{20px}{\centering with faults} & \multicolumn{2}{c|}{PN $\rightarrow$ PN*} & \multicolumn{2}{c|}{PN* $\rightarrow$ BP} & \multicolumn{2}{c}{PN $\rightarrow$ BP} \\
$n$ & & $S$ & $T$ & $S$ & $T$ & $S$ & $T$ \\\midrule
1 &  & 0.40 & 0.38 & 0.17 & 0.25 & 0.50 & 0.54\\
1 & \checkmark & 0.35 & 0.40 & 0.00 & 0.00 & 0.35 & 0.40\\[5px]
2 &  & 0.55 & 0.53 & 0.65 & 0.70 & 0.84 & 0.86\\
2 & \checkmark & 0.40 & 0.43 & 0.54 & 0.56 & 0.73 & 0.75\\[5px]
3 &  & 0.65 & 0.65 & 0.79 & 0.81 & 0.93 & 0.93\\
3 & \checkmark & 0.47 & 0.51 & 0.75 & 0.76 & 0.87 & 0.88\\[5px]
4 &  & 0.65 & 0.65 & 0.89 & 0.89 & 0.96 & 0.96\\
4 & \checkmark & 0.23 & 0.29 & 0.90 & 0.90 & 0.92 & 0.93\\[5px]
5 &  & 0.49 & 0.50 & 0.95 & 0.95 & 0.98 & 0.98\\
5 & \checkmark & 0.25 & 0.32 & 0.94 & 0.94 & 0.95 & 0.96\\[5px]
\midrule
\multicolumn{2}{c|}{\textbf{average}} & \textbf{0.44} & \textbf{0.47} & \textbf{0.66} & \textbf{0.68} & \textbf{0.80} & \textbf{0.82} \\
\multicolumn{2}{c|}{\textbf{median}} & \textbf{0.44} & \textbf{0.47} & \textbf{0.77} & \textbf{0.79} & \textbf{0.90} & \textbf{0.91}\\
 \bottomrule
\end{tabular}
}
\end{table*}

\begin{table}
\centering
\caption{A comparison of the traces of a maximal length of sixteen for both approaches. The PN model is omitted since it has the same traces as the PN* model, given that the helper events are removed from the traces.}
\label{tab:lc-paths}
\begin{tabular}{cc|rrr}
\toprule
 & \multirow{2}{20px}[-2px]{\centering with faults} & \multicolumn{3}{c}{\# of traces}\\[2px]
 $n$ & & \multicolumn{1}{c}{BP only} & \multicolumn{1}{c}{PN* only} & \multicolumn{1}{c}{both}\\\midrule
 1 & & 0 & 78 & 67 \\
1 & \checkmark & 52,718 & 20,730 & 12,740 \\[5px]
2 & & 0 & 305,568 & 148,689 \\
2 & \checkmark & 117,240,548 & 109,323,068 & 30,566,079 \\[5px]
3 & & 0 & 52,604,886 & 44,311,111 \\ 
\bottomrule
\end{tabular}
\end{table}

In \autoref{tab:lc-paths}, we evaluate the equivalency of the models to compare the possible traces of each model (as described in \autoref{sec:comparing-bp-and-pn:equivalency}). Unsurprisingly, the number of paths is exponentially bound to the size of the state space. Therefore, we generated only traces of a maximal length of sixteen and only for a limited number of railways due to memory limitations. This comparison shed much light on the differences between the models, as we now elaborate.

In \autoref{sec:comparing-bp-and-pn}, we showed that for a single track without fault transitions, the resulting language of the BP model is contained in the resulting language of the PN model (\ie $L_{M_{BP}} \subset L_{M_{PN}}$), meaning that some traces are only possible in the PN model. As \autoref{tab:lc-paths} shows, this phenomenon is relevant for multiple tracks as well. To recall, the PN model allows for a redundant raise and lower actions of the barriers after the train approaches.

The inclusion of fault transition to the models increased the models' differences. In addition to traces unique to the PN model, the introduction of fault transitions resulted in unique runs to the BP model. One example for a trace that is accepted by the BP model only is:
$( Approaching\cdot  FaultEntering\cdot Leaving)^\omega$.
In this case, a train enters the crossing zone before the barriers are lowered and then leaves. A train should be able to leave the crossing zone regardless of the barriers state. To understand why this trace is not possible in the PN model with faults, we recall the mechanism (described in \autoref{sec:comparing-bp-and-pn:faults} for complying with the original requirements, which adds arcs to and from $p_9$. This mechanism caused an unexpected side effect. Another trace that is accepted only by the BP model is:
$( Approaching\cdot Lower\cdot Entering\cdot Leaving\cdot Approaching\cdot FaultRaise\cdot Entering\cdot)^\omega$. Here, the barriers were raised, although the system was unaware of this event. Then a train entered the crossing zone. While such behavior is reasonable and may happen, this trace cannot happen in the PN model, revealing another side effect of the mechanism.


\subsection{Additional Petri-Net Models}
We now turn to demonstrate our claims on two other domains to support our hypothesis that the problem is rooted in the language constructs.

\subsubsection{The Dining Philosophers}
\begin{figure*}
\centering
\begin{tikzpicture}
    \pic {dinning_philosophers};
\end{tikzpicture}
\caption{The PN model of~\cite{workcraftDinning} for the dining philosophers problem.}
\label{fig:dining-philosophers}
\end{figure*}
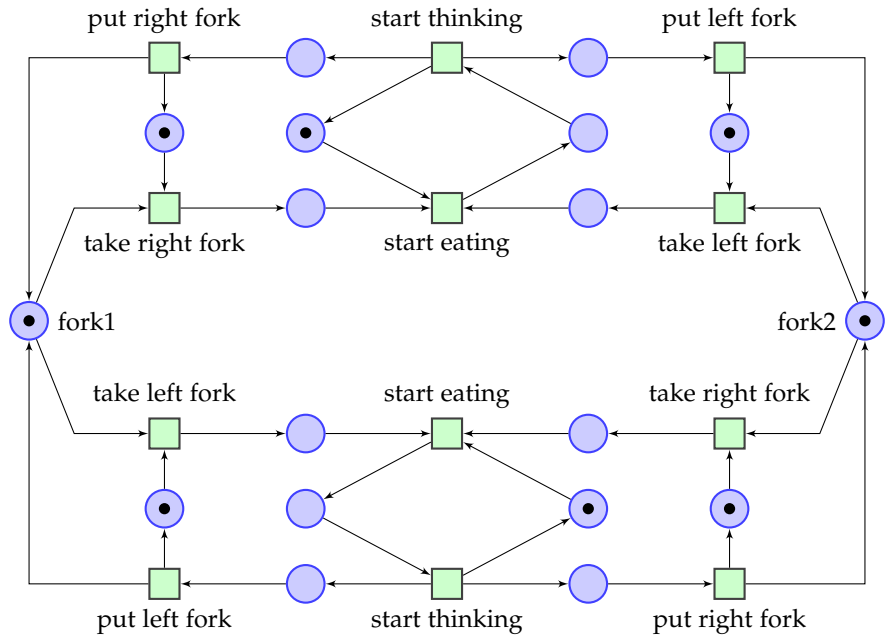

The famous dining philosophers problem has been modeled using PN in many papers and was also modeled in BP~\cite{harel2011model}. In \autoref{lst:dining}, we present the BP implementation for this problem and the PN model in \autoref{fig:dining-philosophers}. We took the PN model from a tutorial for Workcraft --- a framework for interpreted graph models, supporting modeling, verification, and synthesizing such models~\cite{workcraftDinning, poliakov2009workcraft, sokolov2016workcraft}. Both of the implementations define the two basic behaviors of the system, one for philosophers and one for the forks. 

\begin{lstlisting}[
  style=BPjs,
  float=t,
  numbers=none,
  breaklines=true,
  label={lst:dining},
  caption={A behavioral program of the dining philosopher, taken from~\cite{harel2011model}.},
]
const PHIL_COUNT = 2

const ForkTaken = i => 
  [Take(i, "R"), Take((i % PHIL_COUNT) + 1, "L")]
const ForkPut = i => 
  [Put(i, "R"), Put((i % PHIL_COUNT) + 1, "L")]

for (let c = 1; c <= PHIL_COUNT; c++) {
  let i = c
  bthread('Fork ' + i + ' behavior', function () {
    while (true) {
      sync({waitFor: ForkTaken(i), 
              block: ForkPut(i)})
      sync({waitFor: ForkPut(i), 
              block: ForkTaken(i)})
    }
  })

  bthread('Philosopher ' + i + ' behavior', 
      function () {
    while (true) {
      sync({request: [Take(i, 'R'),Take(i, 'L')]})
      sync({request: [Take(i, 'R'),Take(i, 'L')]})
      sync({request: [Put(i, 'R'),Put(i, 'L')]})
      sync({request: [Put(i, 'R'),Put(i, 'L')]})
    }
  })
}
\end{lstlisting}

\begin{lstlisting}[
  style=BPjs,
  numbers=none,
  float=tbph,
  label={lst:dining:liveness},
  caption={Liveness requirements for the dining\newline philosophers. A hot synchronization point specifies that whenever the b-thread arrives to this point, it will eventually proceed.},]
for (let c = 1; c <= PHIL_COUNT; c++) {
  let i = c

  // A taken fork will eventually be released
  bthread('[](take -> <>put)', function () {
    while (true) {
      sync({waitFor: ForkTaken(i)})
      hot(true).sync({waitFor: ForkPut(i)})
    }
  })

  // A hungry philosopher will eventually eat
  bthread('NoStarvation', function () {
    while (true) {
      hot(true).sync({waitFor: 
         [Take(i, 'R'), Take(i, 'L')]})
      hot(true).sync({waitFor: 
         [Take(i, 'R'), Take(i, 'L')]})
      sync({waitFor: [Put(i, 'R'), Put(i, 'L')]})
      sync({waitFor: [Put(i, 'R'), Put(i, 'L')]})
    }
  })
}
\end{lstlisting}

Both implementations may cause the same two problems --- a deadlock and starvation. These problems can be defined as two additional liveness requirements: 1) A picked-up forked will eventually be put down, and; 2) a hungry philosopher will eventually eat. While both PN and BP have tools for detecting liveness problems, BP allows developers to directly specify the liveness requirements, as presented in~\autoref{lst:dining:liveness}. A hot synchronization point specifies that whenever the b-thread arrives  at this point, it will eventually proceed (\ie one of the requested or waited-for events will be selected). 

BP offers several approaches for automatically enforcing a correct execution in terms of liveness, including synthesis, runtime look-ahead, and reinforcement learning~\cite{Yaacov2021Thesis}.
Another option to enforce the execution correctness is explicitly implementing known solutions, like resource ordering and a central arbitrator. Resource ordering in BP can be achieved using priorities. Here, the priority of the philosophers' requests is inversely proportional to their index. We demonstrate the use of priorities in the following example (see \autoref{lst:ttt}). The central arbitrator solution is presented in \autoref{lst:dining:arbitrator}, where philosophers must get a hold on a central semaphore to eat (\ie take and put the forks). The literature also offers PN solutions~\cite{davidrajuh2014verifying}, which we do not present for brevity. Nevertheless, like our previous examples, they require the mechanism specification for interweaving all of the requirements together.

\begin{lstlisting}[
  style=BPjs,
  numbers=none,
  float=tbph,
  label={lst:dining:arbitrator},
  caption={An arbitrator solution to the deadlock requirement.}]
bthread('Semaphore', function () {
  while (true) {
    sync({waitFor: AnyTakeSemaphore})
    sync({waitFor: AnyReleaseSemaphore, 
            block: AnyTakeSemaphore})
  }
})

for (let c = 1; c <= PHIL_COUNT; c++) {
  let i = c
  bthread('Take semaphore ' + i, function () {
    while (true) {
      sync({request: TakeSemaphore(i), 
              block: [Take(i, 'R'),Take(i, 'L')]})
      sync({waitFor: [Put(i, 'R'), Put(i, 'L')]})
      sync({waitFor: [Put(i, 'R'), Put(i, 'L')]})
      sync({request: ReleaseSemaphore(i), 
              block: [Take(i, 'R'),Take(i, 'L')]})
    }
  })
}
\end{lstlisting}

\vspace{2em}
\subsubsection{Tic-Tac-Toe}

\begin{figure*}
\begin{minipage}[b]{0.37\textwidth}
\centering
\begin{lstlisting}[
%   float=thbp,
%   frame=tlrb,
  label={lst:ttt},
  numbers=left,
  caption={A behavioral program for the game of Tic-Tac-Toe, taken from~\cite{harel2010programming}. Lines 1-20 specify the same behavior as the PN model of \autoref{fig:ttt}, and the b-thread in line 29, with l=``first row", specifies the same behavior as the PN model of \autoref{fig:ttt-xwin}.},
]
cells.forEach(c => {
  bthread('Cells cannot be marked ' + 
      'twice', function(){
    sync({waitFor: [O(c), X(c)]})
    sync({  block: [O(c), X(c)]})
  }) })
  
bthread('Enforce turns', function(){
  while(true) {
    sync({waitFor: AnyX, 
            block: AnyO})
    sync({waitFor: AnyO, 
            block: AnyX})
  }
})

bthread('Play randomly', function(){
  while(true)
    sync({request: XOmoves})
})
// ################################
// Above: the behavior of $\autoref{fig:ttt}$.
// ################################

lines.forEach(l => {
  // This b-thread, with l=first row,
  // is equivalent to the extension 
  // in $\autoref{fig:ttt-xwin}$.
  bthread('Detect X win', function(){
    for(let i = 0; i < 3; i++)
      sync({waitFor: 
        [X(l.c1),X(l.c2),X(l.c3)]})
    sync({request: XWin}, 100)
    sync({  block: bp.all })
  })
  
  bthread('Detect O win', function(){
    for(let i = 0; i < 3; i++)
      sync({waitFor:
        [O(l.c1),O(l.c2),O(l.c3)]})
    sync({request: OWin}, 100)
    sync({  block: bp.all })
  })
})

bthread('Detect a tie', function(){
  for(let i = 0; i < 9; i++)
    sync({request: Tie}, 90)
    sync({  block: bp.all })
})
\end{lstlisting}
\end{minipage}
\quad
\begin{minipage}[b]{0.6\textwidth}
\centering
    {\adjustbox{scale=0.9,valign=b}{
    \begin{tikzpicture}
        \pic{ttt};
    \end{tikzpicture}}}
    \caption{The PN model of~\cite{PetriNetDSLblahchain} for the game of Tic-Tac-Toe. The model implements only two requirements out of four.}
    \label{fig:ttt}\vspace{13ex}
  
    {\adjustbox{scale=0.9,valign=b}{
    \begin{tikzpicture}
        \pic[scale=0.6]{tttx};
    \end{tikzpicture}}}
    \caption{Our extended version of the PN model of~\cite{PetriNetDSLblahchain}. The model includes a mechanism for terminating the game when X wins by taking the first row.}
    \label{fig:ttt-xwin}\vspace{6em}\vspace{-2pt}
\end{minipage}
\end{figure*}

The last domain we present is the game of Tic-Tac-Toe. The game is of particular interest for us as it was developed using BP in one of the earliest papers of the paradigm~\cite{harel2010programming}. The requirements of the game are well known and the BP implementation of the game has been published long before this paper. Thus, this domain stands as a touchstone for our hypothesis.

We take the PN model from~\cite{PetriNetDSLblahchain}, which was used to create a domain-specific language (DSL) for the game. As depicted in \autoref{fig:ttt}, this PN model only specifies two requirements of the game: 1) a cell can be marked only once, and; 2) turns --- X and O play in turns where X starts.
The mechanism for integrating these requirements is specified without the use of helper events and is fairly understandable. Nevertheless, there are two additional requirements: 3) The first player to get three marks in a line is the winner, ending the game, and; 4) if no player has won and all nine squares are marked, then the game is over with a tie. Although according to this model, the game ends upon the marking of the last cell, there is no tie declaration (\ie transition). This is where the model gets complicated. Since~\cite{PetriNetDSLblahchain} did not create a complete model of the game, we took this mission on ourselves. In \autoref{fig:ttt-xwin}, we added an interlock mechanism for handling the case that X wins by placing Xs in the first row. The ``game token'' is added to ensure that the game will stop once the token is gone (\ie the game ends). The place ``row$_0$ X counter'' waits for three ``X'' tokens to arrive and then fires them to the ``row$_0$ X win'' transition, together with the ``game token''. Since the transition has no outgoing edges, it acts as a sink that terminates the game. In total, we added 23 edges, two places, one transition, and one token. Since the new model was extremely noisy and hard to understand, we decreased the opacity of the original specification. Handling the other seven lines will require additional 35 edges, seven transitions, and seven places. Adding a tie event with a lower priority than a winning event requires an additional interlocking mechanism.

\autoref{lst:ttt} presents a BP implementation of the game, taken from~\cite{harel2010programming}. Each b-thread represents a single aspect of the game and is unaware of other aspects. For example, the last b-thread repeatedly asks for placing X and O at any of the nine cells. It is unaware of other rules, like turns, which are enforced by the second b-thread. Cell and line b-threads are duplicated for each cell/line. There is one exception to the separation of concerns between the b-threads --- the 90 and 100 numbers in the winning b-threads. These numbers represent the priority of the event. If a game has both nine moves and the last mark wins the game for the X player, then the `XWin' event overcomes the `Tie' event. Notably, the Tie requirement refers to the winning requirement, thus technically, the separation of concerns is violated in the requirements, and the alignment is kept. Nevertheless, there are several solutions to this issue (\eg using context~\cite{Elyasaf2020COBP}), though they are out of the scope of this paper.

Notably, the b-threads in lines 1-20 specify the same behavior as the PN model in \autoref{fig:ttt}. Similarly, the b-thread in line 29, with l=``first row", specifies the same behavior as the PN model in \autoref{fig:ttt-xwin}. This comparison emphasizes the conciseness of the BP model compared to the PN model.

\section{Related Work}
\label{sec:related-work}


Giua and Silva~\cite{giua2017modeling} pointed out that while the use of PNs with state specifications is a very mature area, their use in the design of systems from general behavioral specifications has not been equally successful. The latter issue can in some cases lead to incorrect specifications, faulty implementations, and inconsistent system behavior. Finding a more general approach to system modeling is still an open problem driving several developments in the PN field.

The ability to structurally define the entire system behavior as a function of the behavior of its subsystems is a key factor in designing systems from a general behavior description. Hence there has been much research concentrating on PNs compositionality and sub-PNs interactions representation. Several works introduced a compositional extension of PNs using process algebras~\cite{sobocinski2010representations, sobocinski2013connector, devillers2021articulations}. They provide an approach for a high-level description of interactions, communications, and synchronizations between PNs. In another work, Baldan et al.~\cite{baldan2001compositional} represented PNs interactions by introducing open PN, a generalization of the ordinary model. In open PN, some places, designated as open, reflect interaction with other nets. Concretely, an open place can function as an input or an output (or both), meaning that external PNs can put or remove tokens from it. Kindler and Petrucci~\cite{kindler2009towards} proposed a similar approach that adds an interface for each module, called channel, that specifies the input and the output of the module. All approaches require the definition of an interface between the different components, thus improving their abstraction (as with interfaces of object-oriented programming). Nevertheless, the modelers are still required to consider the mutual dependencies for specifying these interfaces. 
We argue that the BP approach addresses compositionality more naturally, with the ability to compose behaviors without direct consideration of mutual dependencies.

An important part of generalizing PN modeling is the ability to represent an interface of the system with the environment. Plain PNs are not adequate to model systems that can interact with their environment or, in another view, are only partially specified. The above-mentioned open PN~\cite{baldan2001compositional} can also model external interaction, where some nets in the whole model represent the system's environment. In another related work, reactive PN~\cite{eshuis2003reactive} addresses this issue by defining reactive semantics to PN, specifically, splitting the set of transitions into internal and external and modifying its firing rules. These semantics state that if an external transition is enabled, it may fire, while in contrast, internal transition, when enabled, must fire. Such behavior is desired in systems that are specified to react as a consequence of external events. We view the ability to accurately model real-time scenarios in reactive systems to be of great importance. BP approaches an external environment interaction in its semantics~\cite{harel2011behavioral} and implementations~\cite{bar2018bpjs} using the mechanism of super-steps that capture the priority of external events over external ones that reflect the notion of logical execution time~\cite{kirsch2012logical}.



\section{Discussion}
\label{sec:discussion}
There is a qualitative difference between BP models and PN models: BP focuses on breaking systems into requirements, and PN focuses on specifying the components of a system and how they interact. While both approaches have many merits, we argued in this paper that BP is better for specifying system behavior when it is a composition of requirements. To support our claim, we methodologically demonstrated over three problems how PN obliges modelers to specify a mechanism for combining the different behaviors. This results in over-specification, incorrect models, or complicated models that are not directly aligned with the requirements. In view of the long-standing success of Petri nets, we propose, as future work, to get the best of the two by integrating BP semantics (i.e., wait, request, and block) with Petri nets, as previously done with statecharts~\cite{marron2018embedding}. 


The problem of a language that obliges users to say things they do not wish to say does not solely belong to Petri nets. Other programming and modeling languages, compilers, and sometimes even IDEs --- share it. We believe that this problem may be rooted in the relationship between programming languages and the early, ``hard'' version of the linguistic relativity hypothesis. For years, programming languages have directed users to adapt their thinking. Behavioral programming is different in that its primary design goal is to allow its users to specify the system behavior in a natural and intuitive manner that is aligned with how \textit{they perceive} the system requirements. We are not saying that BP does not share this problem; however, the BP community is devoted to refining and extending the paradigm to eliminate these problems. For example, a recent extension to the paradigm~\cite{Elyasaf2020COBP} has pointed out that the absence of context idioms in BP obliges users to define mechanisms for specifying context-dependent requirements. These mechanisms either break the alignment to the requirements or break the correctness of the model. To allow users a more natural specification of their context-dependent requirements, the extension of~\cite{Elyasaf2020COBP} adds context idioms as first-class citizens of the language. 

In practice, software projects rarely start with well-defined requirements. Reasonably, it may be related to the challenge of maintaining the requirement documents, the design documents, and the traceability between the requirements and the code. Generally speaking, requirements are how people describe their system. Thus, a shift left of modeling/programming languages towards a more natural specification/programming of the requirements may lead to an evolutionary step in the field. There are other possible solutions and approaches to this problem. We believe that the software engineering community will benefit from adopting modern linguistics approaches and searching for these solutions.

\ifCLASSOPTIONcompsoc
  \section*{Acknowledgments}
\else
  \section*{Acknowledgment}
\fi
This research was partially supported by grant \#2714/19 from the Israeli Science Foundation and by the Israeli Smart Transportation Research Center (ISTRC). 

The authors would like to thank Assaf Marron for a thorough review of the draft of this paper. 

\ifCLASSOPTIONcaptionsoff
  \newpage
\fi

\bibliographystyle{IEEEtran}
\bibliography{bp_pn}

%

\begin{IEEEbiography}[{\includegraphics[width=1in,height=1.25in,clip,keepaspectratio]{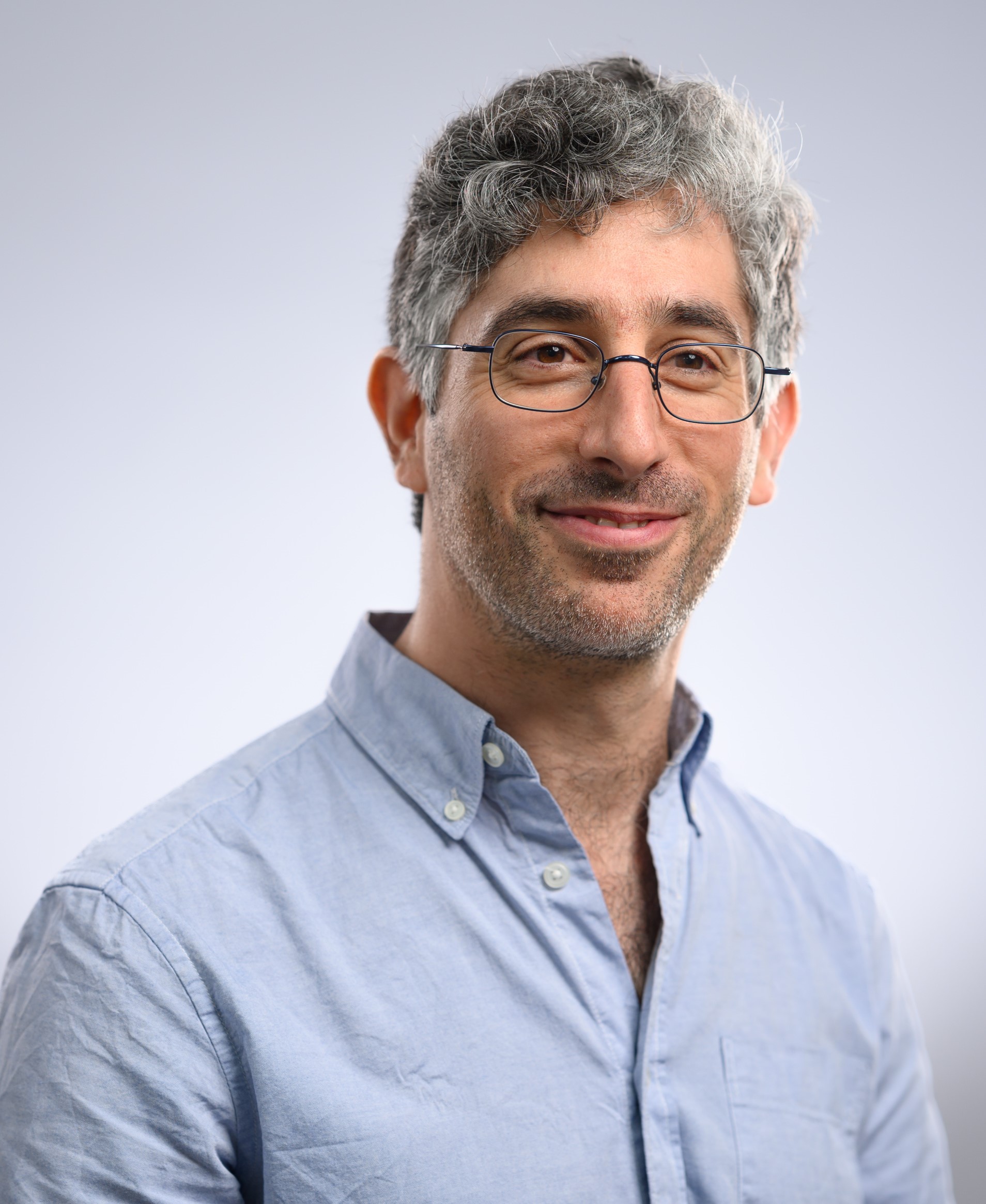}}]{Achiya Elyasaf} is a faculty at the Department of Software and Information Systems Engineering at the Ben-Gurion University. His research areas include software engineering and artificial intelligence. He developed context-oriented behavioral programming, an extension to the paradigm presented in this paper, that adds first-class context idioms for specifying context-dependent behaviors. 

Dr. Elyasaf received his Ph.D. at the Ben-Gurion University under the supervision of Prof. Moshe Sipper. His work was recognized as one of the most important achievements of AI in games, placed next to famous achievements, such as the wins of Deep Blue over Kasparov and Watson in Jeopardy! (N. Bostrom, 2014. ``Superintelligence: Paths, Dangers, Strategies''). He continued as a post-doc at the Weizmann Institute of Science in David Harel's group. In 2021 he co-founded Provengo technologies, which provides software-testing solutions based on behavioral programming.
\end{IEEEbiography}

\begin{IEEEbiography}[{\includegraphics[width=1in,height=1.25in,clip,keepaspectratio]{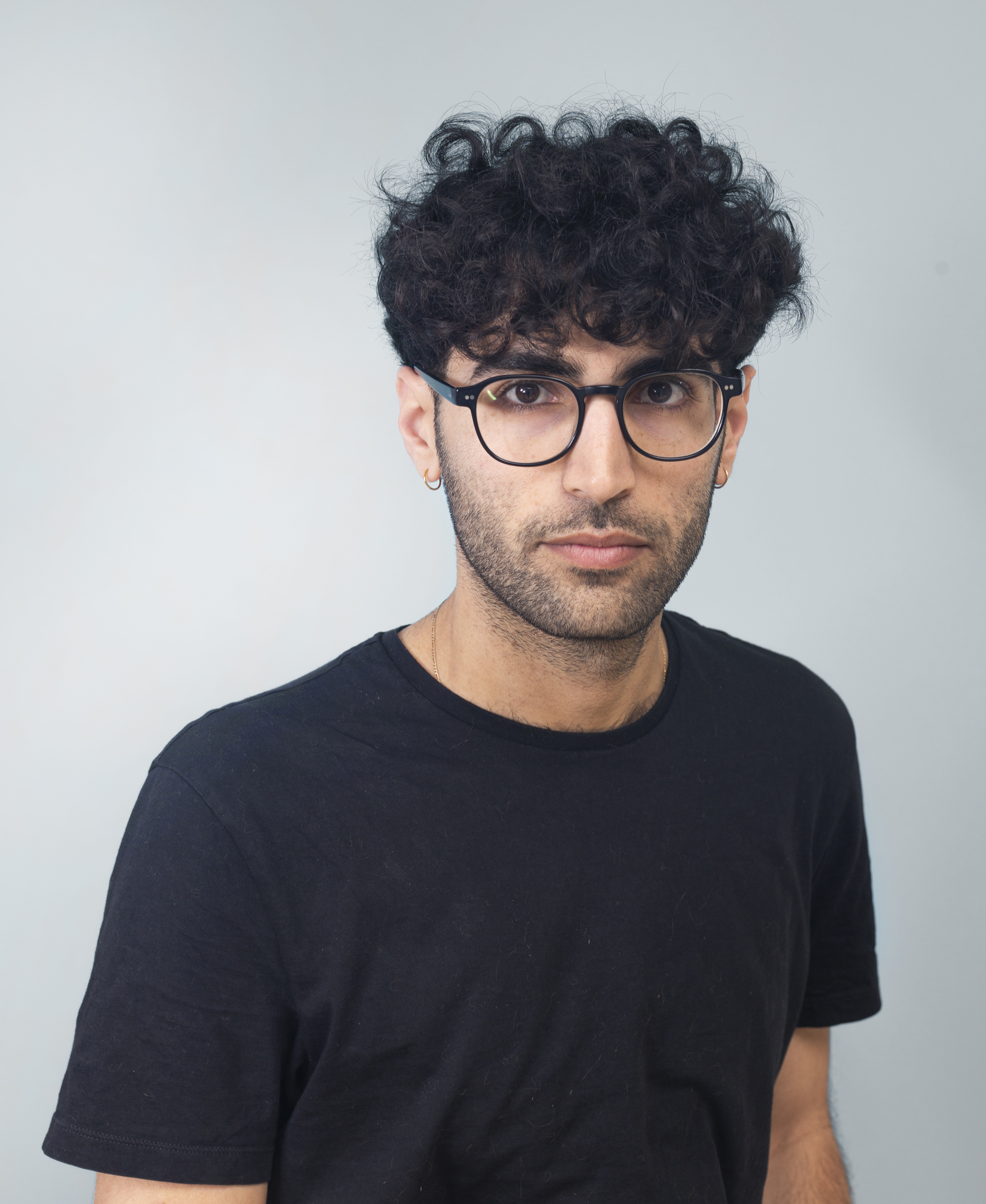}}]{Tom Yaacov}
received his B.Sc. (Magna cum laude) in industrial engineering and management and his M.Sc. (summa cum laude) in computer science from the Ben-Gurion University of the Negev, Beer Sheva, Israel, where he is currently working towards a Ph.D. degree.

His current research involves the application of machine learning algorithms to software engineering practices and disciplines.

Tom received several fellowships, including dean's list.
\end{IEEEbiography}

\vfill

\newpage

\begin{IEEEbiography}[{\includegraphics[width=1in,height=1.25in,clip,keepaspectratio]{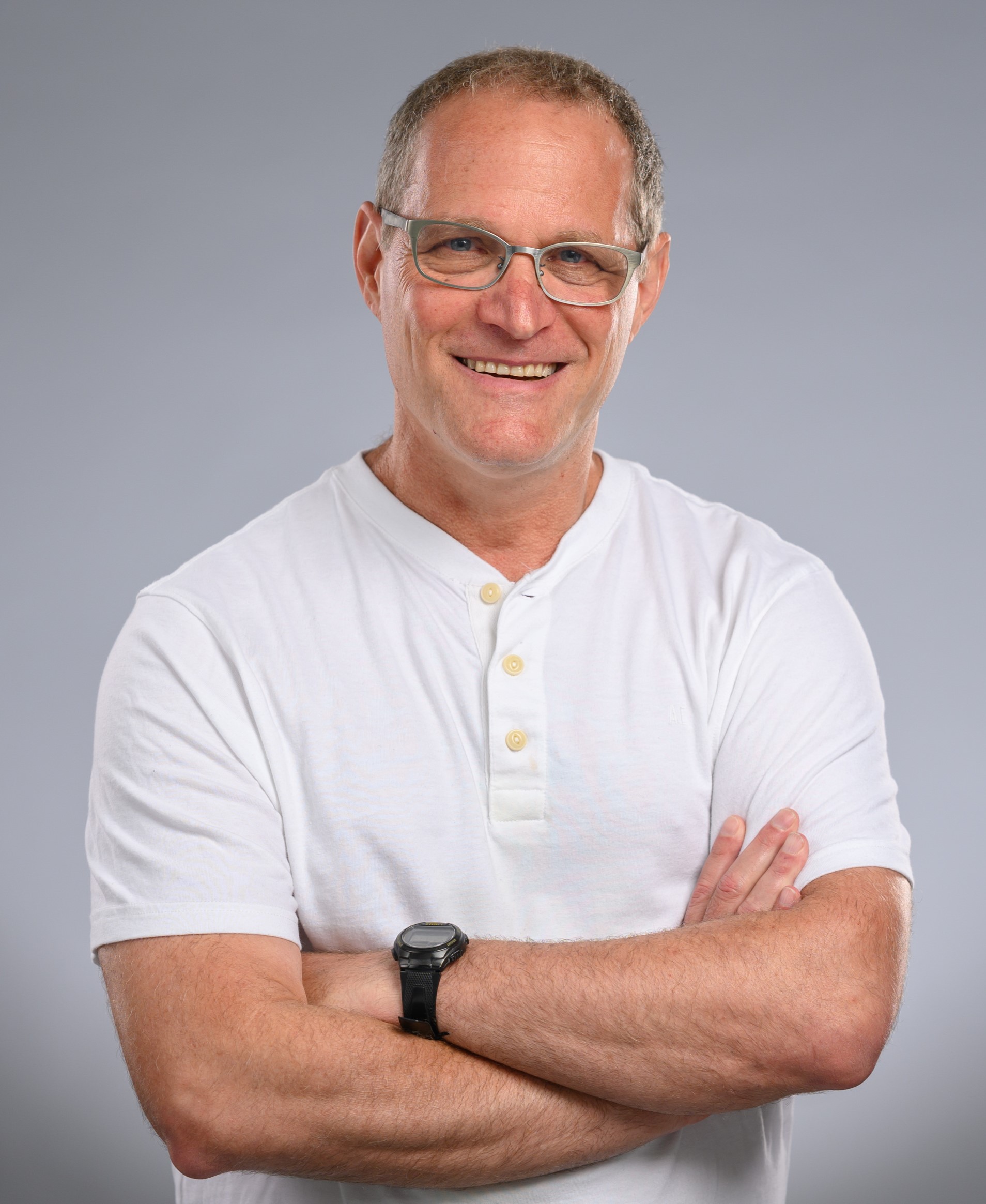}}]{Gera Weiss} is a faculty at the Department of Computer Science at the Ben Gurion University of the Negev. His research areas include software engineering, control theory, and formal methods. He is a co-author of the papers that introduced Behavioral Programming, the method that we compare to Petri Nets in this paper, promoting alignment of code with requirements by allowing negative statements (anti-scenarios) such  as ``don't produce two Xs in a row".

Prof. Weiss obtained his Ph.D. at the Weizmann Institute of Science under the supervision of Prof. Zvi Artstein and Prof. Amir Pnueli. He continued as a post-doc at the University of Pennsylvania in Prof. Rajeev Alur's group and serves as a faculty at the Ben-Gurion University of the Negev since 2010. He now serves as the head of the software engineering program. In 2021 he co-founded Provengo technologies, which provides software-testing solutions based on behavioral programming.
\end{IEEEbiography}



\end{document}